\documentclass[twocolumn]{aastex631}
\usepackage{lineno}
\received{xxx}
\revised{xxx}
\accepted{xxx}

\submitjournal{PASP}

\shorttitle{Measurement accuracy of non-parametric morphological indicator}
\shortauthors{Luo et al.}

\begin{document}

\title{Evaluating the Accuracy of Non-parametric Galaxy Morphological Indicator Measurements in the CSST Imaging Survey}

\correspondingauthor{Jian Ren, Nan Li, F.S. Liu}
\email{renjian@bao.ac.cn; nan.li@nao.cas.cn; fsliu@nao.cas.cn}

\author{Yuchong Luo}\thanks{This author contributed equally to this work}
\affiliation{National Astronomical Observatories, Chinese Academy of Sciences, Beijing 100101, China}
\affiliation{Key Laboratory of Space Astronomy and Technology, National Astronomical Observatories, Chinese Academy of Sciences, 20A Datun Road, Chaoyang District, Beijing 100101, China}

\author{Anhe Sha}\thanks{This author contributed equally to this work}
\affiliation{National Astronomical Observatories, Chinese Academy of Sciences, Beijing 100101, China}


\author{Jian Ren}
\affiliation{National Astronomical Observatories, Chinese Academy of Sciences, Beijing 100101, China}
\affiliation{Key Laboratory of Space Astronomy and Technology, National Astronomical Observatories, Chinese Academy of Sciences, 20A Datun Road, Chaoyang District, Beijing 100101, China}

\author{Xin Zhang}
\affiliation{National Astronomical Observatories, Chinese Academy of Sciences, Beijing 100101, China}
\affiliation{Key Laboratory of Space Astronomy and Technology, National Astronomical Observatories, Chinese Academy of Sciences, 20A Datun Road, Chaoyang District, Beijing 100101, China}

\author{Xianmin Meng}
\affiliation{National Astronomical Observatories, Chinese Academy of Sciences, Beijing 100101, China}

\author{Nan Li}
\affiliation{National Astronomical Observatories, Chinese Academy of Sciences, Beijing 100101, China}
\affiliation{Key Laboratory of Space Astronomy and Technology, National Astronomical Observatories, Chinese Academy of Sciences, 20A Datun Road, Chaoyang District, Beijing 100101, China}
\affiliation{School of Astronomy and Space Science, University of Chinese Academy of Science, Beĳing 100049, China}

\author{F.S. Liu}
\affiliation{National Astronomical Observatories, Chinese Academy of Sciences, Beijing 100101, China}
\affiliation{School of Astronomy and Space Science, University of Chinese Academy of Science, Beĳing 100049, China}
\affiliation{Key Laboratory of Optical Astronomy, National Astronomical Observatories, Chinese Academy of Sciences, 20A Datun Road, Chaoyang District, Beijing 100101, China}



\begin{abstract}

The Chinese Space Station Telescope (CSST) is China's upcoming next-generation ultraviolet and optical survey telescope, with imaging resolution capabilities comparable to the Hubble Space Telescope (HST). In this study, we utilized a comprehensive sample of 3,679 CSST realistic mock galaxies constructed from HST CANDELS/GOODS-North deep imaging observations, with stellar masses $\log\left(M_{*} / M_{\odot}\right) > 9.0$ and redshifts $z < 2$. We evaluate the detection capabilities of CSST surveys and the accuracy in measuring the non-parametric morphological indicators ($C$, $A$, $Gini$, $M_{\rm 20}$, $A_{\rm O}$, $D_{\rm O}$) of galaxies. Our findings show that in terms of galaxy detection capabilities, CSST's deep field surveys can achieve the same level as HST's deep field observations; however, in wide-field surveys, CSST exhibits a significant deficiency in detecting high-redshift, low-mass, low-surface-brightness galaxies. Regarding the measurement of galaxy morphology, CSST's deep field surveys achieve high accuracy across all indicators except for the asymmetry indicator ($A$), whereas its wide-field surveys suffer from significant systematic biases. We thus provide simple correction functions to adjust the non-parametric morphological indicators obtained from CSST's wide-field and deep-field observations, thereby aligning CSST measurements with those from HST. This adjustment enables the direct application of non-parametric morphological classification methods originally developed for HST data to galaxies observed by CSST.

\end{abstract}



\section{Introduction} \label{sect:intro}

Galaxy morphology research is dedicated to the classification and analysis of galaxy shapes and structures, offering a direct illustration of the spatial distribution of stars and dust within galaxies \citep{Conselice2020}. Furthermore, it provides crucial insights into the formation and evolutionary history of galaxies, which been consistently supported and validated by numerous studies \citep[e.g.,][]{Conselice+etal+2008,Mortlock+etal+2013,Conselice+etal+2014,Huertas-Company+etal+2016}. 
This field has long been a central research direction in astronomy, with its progress closely linked to advancements in observational techniques, the evolution of theoretical models, and technological innovations. The approach to classifying galaxies based on their morphological characteristics has a long history, tracing back to the pioneering contributions of Edwin Hubble in the 1920s \citep{Hubble+1926}. Hubble's contributions led to the creation of the famous Hubble sequence, which categorizes galaxies into three primary types —- elliptical, spiral, and irregular —- based on their visual characteristics. Since then, the Hubble sequence has become a fundamental framework in the study of galaxy morphology.

In recent decades, the development of powerful telescopes such as the Hubble Space Telescope (HST) and advanced ground-based telescopes, along with cutting-edge imaging technologies, has enabled detailed observations of galaxies across a broad range of wavelengths. Several large-scale survey projects have been carried out, including the Sloan Digital Sky Survey \citep[SDSS;][]{York+etal+2000}, the Cosmic Evolution Survey \citep[COSMOS;][]{Scoville+etal+2007}, the Cosmic Assembly Near-infrared Deep Extragalactic Legacy Survey \citep[CANDELS;][]{Grogin+etal+2011,Koekemoer+etal+2011}, and the Dark Energy Survey \citep[DES;][]{Abbott+etal+2018}. These surveys have accumulated vast datasets containing rich galaxy images, providing new perspectives on the structures and compositions of galaxies \citep[e.g.,][]{Kauffmann+etal+2003,Baldry+etal+2004,Conselice+2006,Mortlock+etal+2013}. Meanwhile, theoretical models and simulations have also seen significant advancements. As a result, numerous methods for galaxy morphology classification have been proposed. These methods include bulge and disk decomposition \citep{Freeman+1970,Kent+1985}, parametric morphological measurements (e.g.,  S{'e}rsic fitting; \citep{Sersic+1963}), and non-parametric galaxy morphological indicator measurements. These approaches are more objective than manual visual inspection and facilitate direct comparison of results among different researchers.

The non-parametric galaxy morphological indicators method offers a rapid and quantitative tool for analyzing the morphological structures of galaxies. Unlike visual methods (such as the Galaxy Zoo project) \citep{Lintott+etal+2011} and parametric methods, this approach employs a range of indicators to identify various morphological and sub-structural features of galaxies, as shown by \cite{Abraham+etal+1994}, \cite{Conselice+2003}, \cite{Lotz+etal+2004}, \cite{Wen+etal+2014}, \cite{Pawlik+etal+2016}, and \cite{Peth+etal+2016}. Although it can be affected by various observational effects \citep[e.g.,][]{Lotz+etal+2006, Ren+etal+2024, Wang+etal+2024, Yu+etal+2023}, it is especially well-suited for rapid quantitative analysis and the study of morphological structures.

The Chinese Space Station Telescope \citep[CSST;][]{Zhan+2021} is a 2-meter aperture survey space telescope with a field of view exceeding 1.1 square degrees, planned for launch in the coming years. It will conduct multi-band imaging and slitless spectroscopic surveys of approximately 17,500 square degrees of the sky, with 400 square degrees dedicated to deep observations. The CSST imaging observations are equipped with seven filters (NUV, u, g, r, i, z, and y). Its large aperture and wide field of view enable detailed imaging of galaxies across diverse distances and environments, offering a valuable dataset for morphological studies.

Given the upcoming large-scale imaging surveys by the Chinese Space Station Telescope, it is crucial to evaluate its detection capabilities for galaxies. This assessment will help estimate the size of the galaxy sample it will provide and analyze the recovery of non-parametric morphological measurements under both wide-field and deep-field observation modes. In this work, we utilize CSST mock images to evaluate its detection capabilities. Since these mock images are based on deep-field observations from the HST, we use the measurement parameters from HST deep-field images as a reference standard to assess the performance of CSST in recovering non-parametric morphological measurements of galaxies.

This paper is structured as follows: In Section 2, we describe our data and sample. In Section 3, we introduce the non-parametric morphological indicators and outline the methods used for source detection and morphological indicator measurements. In Section 4, we present our results. Discussions are provided in Section 5. Finally, we summarize our conclusions in Section 6. Throughout the paper, we adopt a cosmology with $\Omega_{\rm M} = 0.3$, $\Omega_{\Lambda} = 0.7$, and $H = 70\,\text{km}\,\text{s}^{-1}\,\text{Mpc}^{-1}$. All magnitudes are in the AB system.

\section{Data and sample} \label{dataset}

\subsection{The HST observation imaging data}

\begin{figure*}[htbp]
    \centering
    \includegraphics[width=14cm]{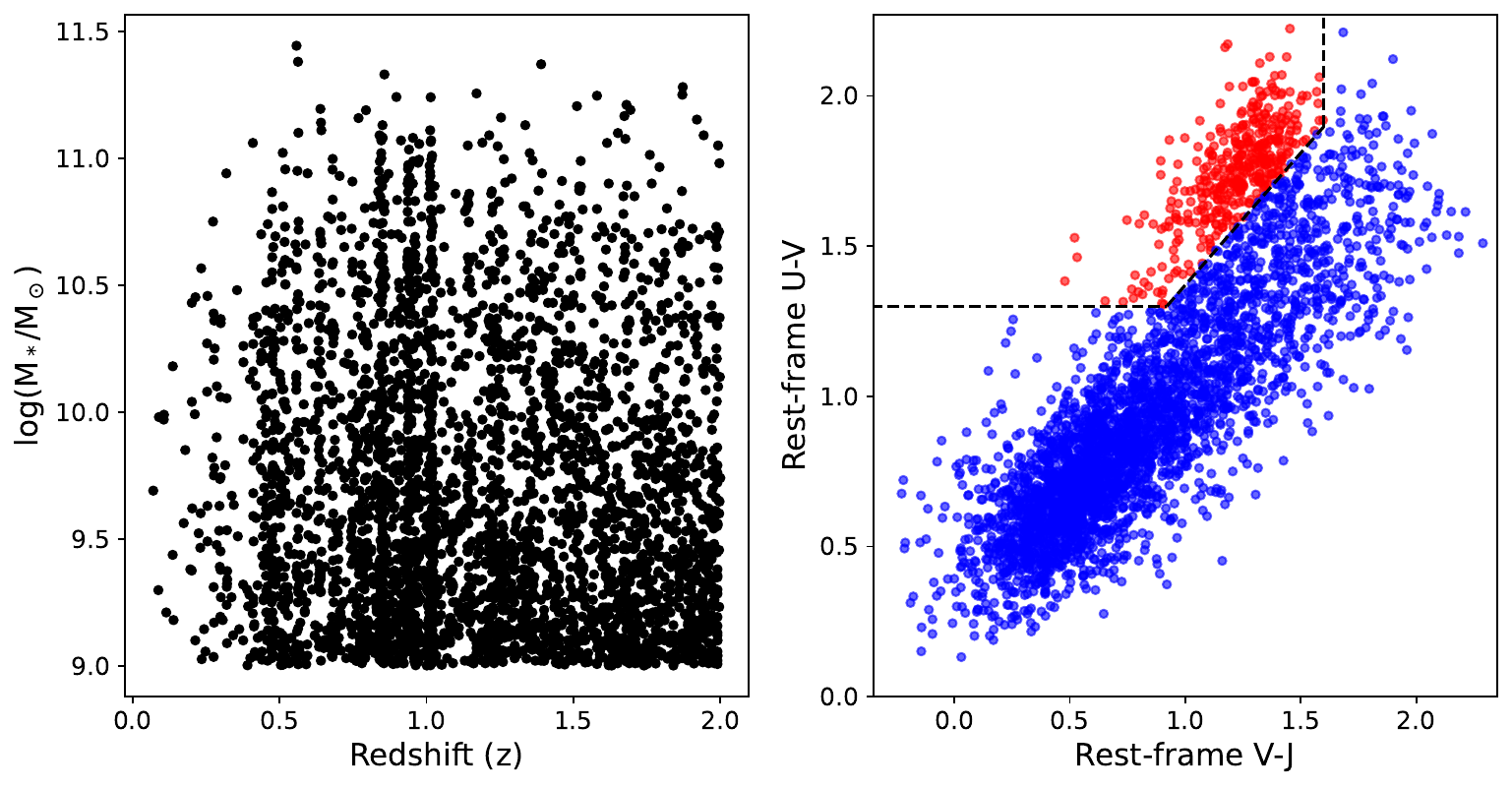}
    \caption{Left panel: mass and redshift distribution of the galaxies in the parent sample. Right panel: rest-frame UVJ diagram of the parent sample. Star-forming galaxies (SFGs) and quiescent galaxies (QGs) were classified using the method proposed by \cite{Williams+etal+2009} and are represented by red and blue markers, respectively.}
    \label{parent_sample_character}
\end{figure*}

In this study, we utilized the V2.5 data release from the Hubble Legacy Fields (HLF) 
project\footnote{https://archive.stsci.edu/prepds/hlf/} for GOODS-North \citep{Illingworth+2019}. 
The observations included in this version of the HLF-GOODS-N field span from July 2002 to November 2018, 
comprising data from 29 distinct HST programs. This dataset includes 5,817 exposures, totaling 3.6 million seconds of observation time. The HLF-GOODS-N dataset incorporates 11 filters: two from WFC3/UVIS (WFC3/UV F275W and F336W), five from ACS/WFC (F435W, F606W, F775W, F814W, and F850LP), and four from WFC3/IR (F105W, F125W, F140W, and F160W). For each filter, the HLF-GOODS-N team provides drizzled science images and weight images. To establish a robust astrometric framework, a global astrometric solution was bootstrapped from the smaller datasets. All the HLF image mosaics have been produced using a coordinate grid tied to an absolute Gaia DR2 reference frame. The primary reason for selecting this field is that, compared to the HLF-GOODS-S field, HLF-GOODS-N includes a larger area of WFC3/UVIS ultraviolet observations, which aligns closely with the ultraviolet observation bands of the Chinese Space Station Telescope (CSST). Additionally, this region provides multi-band high-resolution images spanning from ultraviolet to optical and near-infrared, making it one of the deepest and most comprehensive multi-band observational datasets available to date.

\subsection{Sample selection}

We selected our initial sample from the GOODS-North photometric catalog \citep{Barro+etal+2019}. The sample selection criteria are as follows:

1. Photflag=0 to ensure that the photometry is reliable and spurious sources are removed.

2. CLASS\_STAR$\leq$0.9 to reduce contamination by stars.

3. GALFIT $f_{\rm H}=0$ to select galaxies with relatively regular morphology.

4. Redshifts $z < 2$ and stellar mass $\log\left(\rm M_{*} / M_{\odot}\right) > 9.0$

5. Multi-band images are available and not contaminated by the borders of the mosaic, 
bad pixels, asterisms, or bright sources.

After applying the above cuts, we selected 3,679 galaxies. 
The left panel of Figure~\ref{parent_sample_character} shows the distribution of stellar mass and redshift for the selected sample. As can be seen from the figure, the sample is relatively complete within the specified redshift and stellar mass ranges. 
The right panel of Figure~\ref{parent_sample_character} presents the UVJ diagram for the selected galaxies, which we classify into star-forming and quiescent populations based on the method of \cite{Williams+etal+2009}. The figure shows that the selected galaxies exhibit distinct characteristics in the UVJ diagram, consistent with the properties of stellar populations. It should be noted that, in our sample selection, we only included galaxies with good photometry and reliable \texttt{GALFIT} fitting results, which means that more irregular galaxies—those for which non-parametric morphological indicators are particularly useful—are not explicitly evaluated in this work. This is because, during the mock imaging process, we need to accurately determine the properties of each sample galaxy. For galaxies that are in strong interaction or merging stages, current deblending techniques are not capable of achieving effective separation. Therefore, our sample does not include galaxies in the merging process \citep{Fang+etal+2018}. We proceed with caution, clearly stating this limitation in sample selection and alerting readers to its potential impact.

\subsection{The CSST mock imaging data}

The CSST simulated image data encompasses all 7 bands: NUV, u, g, r, i, z, and y. The central positions and coverage areas of the simulated images correspond to those of the HST observed images. For each band, there are 10 simulations with exposure times ranging from $150\rm s \times 1$ to $250\rm s \times 8$. Below is a brief introduction to the process of creating the simulated images:

\begin{figure*}
        \centering
        \includegraphics[width=14cm, angle=0]{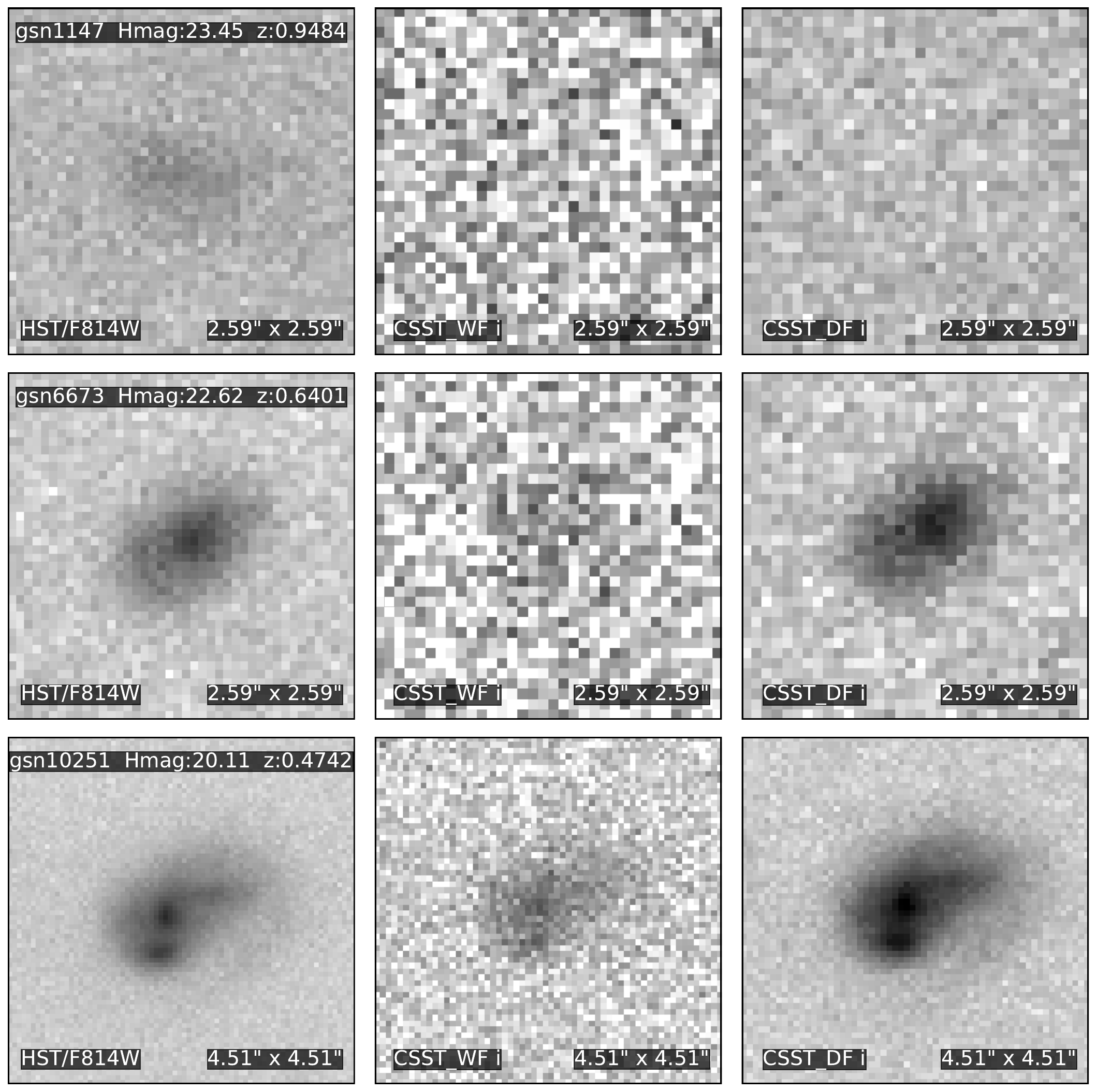}
        \caption{The examples of galaxy images are organized with each column, from left to right, showing the HST image, the corresponding CSST wide field simulated image, and the corresponding CSST deep field simulated image. Each row, from top to bottom, represents a different detection scenario: a galaxy detected only in the HST image, a galaxy detected in both the HST image and the CSST's deep-field image, and a galaxy detected in all three images—the HST, CSST's wide-field, and CSST's deep-field images. Each pixel’s flux within the image has been normalized.}
        \label{example}
\end{figure*}

In the GOODS-North region, the selected sample galaxies exhibit high-quality imaging. These data include HST image data across 10 bands: F275W, F435W, F606W, F775W, F814W, F850LP, F105W, F125W, F140W, and F160W. We utilized unsaturated, high signal-to-noise ratio stars within the field of view to construct the point spread function (PSF) for each band. Subsequently, we built the kernel function between the PSFs of all bands and the F160w band. By performing PSF matching on the images of all bands to the F160w band, we ensured that the same region of a galaxy corresponds to the same physical area across different band images. Centered on the galaxy's core, several radii are determined based on a power-law distribution according to the galaxy's size. The galaxy is divided into multiple rings based on its ellipticity, position angle, and the aforementioned radii. For each annular region of the galaxy, multi-band image data are selected to calculate the photometric data of each ring. Using the \texttt{LePhare}\footnote{http://lephare.lam.fr} software \citep{Arnouts+etal+1999, Ilbert+etal+2006} and the multi-band photometric data of each ring, the corresponding SEDs are fitted. This allows for the derivation of fitted spectral data for different regions of the galaxy.

\renewcommand{\arraystretch}{1.1}  
\begin{deluxetable*}{cccccccc}
\tablenum{1}
\tablecaption{The corresponding relationship between the simulated galaxy morphology in each band of CSST and the HST image.\label{Tab1}}
\tablewidth{0pt}
\tablehead{
\colhead{CSST} & \colhead{NUV} & \colhead{u} & \colhead{g} & \colhead{r} & \colhead{i} & \colhead{z} & \colhead{y} \\[-17pt]
}
\startdata
HST & F275W, F435W & F435W & F435W & F606W & F814W & F850LP & F850LP \\
\enddata
\end{deluxetable*}

The morphology of galaxies under CSST observation conditions is interpolated based on actual HST observational data. For the 7 bands of CSST imaging, corresponding HST data are used as the morphological input into the simulation system. The correspondence is shown in Table~\ref{Tab1}. The morphology of the galaxy and the SEDs of its corresponding components enable the simulation of CSST band galaxy imaging data. 
The deep field observation images of the HST have higher spatial resolution than those of the CSST, and are deeper than both the wide-field and deep-field survey images of CSST. To simulate the observational conditions of CSST, we resampled the HST images and degraded the signal-to-noise ratio (S/N) to match the expected S/N of CSST.
These data represent ideal conditions, to which CSST instrument noise must be added, including detector readout noise of 5 e-/pixel, dark current of 0.02 e-/s/pixel, and photon noise from the target source. Additionally, sky background is incorporated into the simulated images. The background is calculated based on the average background value of HST in the corresponding CSST band, measured in pixels. For the NUV, u, g, r, i, z, and y bands, the sky background values are [0.00261, 0.01823, 0.15897, 0.20705, 0.21433, 0.12658, 0.03755] e-/s/pixel. These simulated images and a ReadMe file are available on GitLab\footnote{https://csst-tb.bao.ac.cn/code/zhangxin/realisticgalsim}. The exposure times for the CSST wide and deep fields are listed in Table~\ref{Tab2}. In Figure~\ref{example}, we present some original HST images alongside their corresponding CSST wide-field and deep-field simulated images.

\renewcommand{\arraystretch}{0.95}
\begin{deluxetable}{ccc}
\tablenum{2}
\tablecaption{Exposure times(s) for different fields and bands of CSST.\label{Tab2}}
\tablewidth{0pt}
\tablehead{
\colhead{Band} & \colhead{Wide Field} & \colhead{Deep Field} \\[-15pt]
}
\startdata
u,g,r,i,z  &  300 & 2000  \\
NUV,y  & 600    &   4000  \\
\enddata
\end{deluxetable}

\section{methods} \label{methodology}

\subsection{The non-parametric morphological indicators}

\subsubsection{$C$ and $A$}

C: Light concentration, which describes the degree of light concentration in a galaxy image, is commonly defined as \citep[e.g.,][]{Abraham+etal+1994, Bershady+etal+2000, Conselice+2003} :

\begin{equation}
C=5 \times \log \left(\frac{\mathrm{R}_{\text {outer }}}{\mathrm{R}_{\text {inner }}}\right)
\end{equation}

In this formula, $R_{\text{outer}}$ and $R_{\text{inner}}$ refer to the radii that encompass 80\% and 20\% (in this paper) of the total flux of a galaxy, respectively.

A: The asymmetric indicator is defined as the difference between the rotated image and the original galaxy image. It is calculated by subtracting the image rotated by 180 degrees from the original image \citep{Conselice+etal+2000, Conselice+2003} and then subtracting the background noise. In this paper, we correct this bias according to the formula proposed by \cite{Wen+2016}, and modify the definition of A as follows:

\begin{equation}
A=\frac{\sum\left|I_0-I_{180}\right|-\delta_2}{\sum\left|I_0\right|-\delta_1}
\end{equation}

In this formula, $\delta_1=f_1 \times \Sigma\left|B_0\right|$, $f_1=\mathrm{N}_{\text {flux }<1 \sigma} / \mathrm{N}_{\text {all }}$, 
$\delta_2=f_2 \times \Sigma \left|B_{0}-B_{180} \right|$, $f_2=\mathrm{N}_{|\text{flux}|<\sqrt{2}\sigma} / \mathrm{N}_{\text {all }}^{\prime}$.

The two correction factors, $\delta_1$ and $\delta_2$, respectively represent the noise contributions to the flux image $I_0$ and the residual image $I_{180}$, where $B_{0}$ and $B_{180}$ represent the corresponding backgrounds of $I_0$ and $I_{180}$, respectively. The fraction of pixel counts in the galaxy source that are dominated by noise is denoted by $f_1$, and $f_2$ represents the fraction of pixel counts in the residual image that are dominated by noise. The total number of pixels in the source and residual images is denoted as $N_{\text {all}}$ and $N_{\text {all}}^{\prime}$, respectively. The standard deviation of noise in $I_0$ is represented by $\sigma$.

\subsubsection{$Gini$ and $M_{\rm 20}$}

The $Gini$ indicator is a statistical tool used in economics to quantify wealth inequality. Some studies have adapted this indicator to investigate the light distribution across pixels in galaxy images \citep{Lotz+etal+2004}. It can be calculated as follows,

\begin{equation}
G=\frac{1}{|\bar{f}| n(n-1)} \sum_i^n(2 i-n-1)\left|f_i\right| 
\end{equation}

In this formula, $\bar{f}$ represents the average flux per pixel, n represents the number of pixels in the galaxy image, and i ranges from 0 to n. $f_{i}$ represents the flux of the $i$-th pixel when sorted in ascending order. A higher $Gini$ value indicates a more uneven distribution of light in the galaxy, with $Gini$ = 1 meaning all the light is concentrated in a single pixel. A lower Gini value indicates a more uniform distribution of light, with $Gini$ = 0 meaning all pixels have the same flux $\bar{f}$. \cite{Lisker+2008} proposed that the measurement of the $Gini$ coefficient is related to the signal-to-noise ratio (SNR) of the image.

The second-order moment indicator, $M_{\rm 20}$, measures the deviation of the brightest 20\% of pixels from the "center" of the galaxy image, making it useful for identifying brighter structures. To measure $M_{\rm 20}$, all pixels in the galaxy image above a certain threshold are sorted in descending order based on their flux. The normalized spatial second-order moment of the brightest 20\% of these pixels, relative to the galaxy's center, is then calculated. Similarly, \cite{Lotz+etal+2004} formulated the following expression:

\begin{equation}
M_{20}=\log \left(\frac{\sum_i M_i}{M_{\text {tot }}}\right), \sum f_i<0.2 \times f_{\text {tot }}
\end{equation}

\begin{equation}
M_{\mathrm{tot}}=\sum_i^n M_i=\sum_i^n f_i\left[\left(x_i-x_m\right)^2+\left(y_i-y_m\right)^2\right]
\end{equation}

In these equations, $f_{i}$ represents the flux value of the $i$-th pixel in the segmented map, arranged in descending order. ($x_{m}$, $y_{m}$) denotes the centroid pixel that minimizes the total moment $M_{\mathrm{tot}}$.

\subsubsection{$A_{\rm O}$ and $D_{\rm O}$}

The $A$ indicator is limited in its ability to detect asymmetric structures in the outer regions of galaxies. To overcome these limitations, \cite{Wen+etal+2014} introduced a new non-parametric method, $A_{\rm O}$-$D_{\rm O}$, specifically designed for detecting asymmetries in the outer regions of galaxies. This method divides the galaxy image into two regions, segmented by isophotes, which are apertures that enclose half of the total galaxy light. The inner region, referred to as the inner half-light region (IHR), typically encompasses a smaller area but has higher surface brightness compared to the outer region, known as the outer half-light region (OHR).

The formula for $A_{\rm O}$ is as follows:

\begin{equation}
A_{\mathrm{O}}=\frac{\sum\left|I_{\mathrm{O}}-I_{\mathrm{O}}^{180}\right|-\delta_2}{\sum\left|I_{\mathrm{O}}\right|-\delta_1}
\end{equation}

Similar to $A$, the correction factors are defined as $\delta_1=f_1 \times \Sigma\left|B_{\mathrm{O}}\right|$, $f_1=\mathrm{N}_{\text {flux }<1 \sigma} / \mathrm{N}_{\text {all }}$, 
$\delta_2=f_2 \times \Sigma \left|B_{\mathrm{O}}-B_{\mathrm{O}}^{180} \right|$, $f_2=\mathrm{N}_{|\text{flux}|<\sqrt{2}\sigma} / \mathrm{N}_{\text {all }}^{\prime}$. $I_{\mathrm{O}}$ and $I_{\mathrm{O}}^{180}$ refer to the OHR and 180$^\circ$-rotated OHR images. In these expressions, $\delta_1$ and $\delta_2$ represent the noise contributions to the flux image $I_{\mathrm{O}}$ and the residual image $I_{\mathrm{O}}^{180}$, respectively. Similarly, $B_{\mathrm{O}}$ and $B_{\mathrm{O}}^{180}$ represent the corresponding backgrounds of $I_{\mathrm{O}}$ and $I_{\mathrm{O}}^{180}$, respectively. $f_1$ and $f_2$ are the number fractions of pixels dominated by noise in the galaxy and residual images.

The formula for  $D_{\rm O}$ is as follows:

\begin{equation}
D_{\mathrm{O}}=\frac{\sqrt{\left(x_{\mathrm{O}}-x_{\mathrm{I}}\right)^2+\left(y_{\mathrm{O}}-y_{\mathrm{I}}\right)^2}}{R_{\mathrm{eff}}}
\end{equation}

In this equation, ($x_{\mathrm{I}}$, $y_{\mathrm{I}}$) and ($x_{\mathrm{O}}$, $y_{\mathrm{O}}$) represent the centroid coordinates of the IHR and OHR, respectively, while $R_{\mathrm{eff}}$ denotes the normalized effective radius of the aperture.

\subsection{Source detection and parameter measurements} \label{measurements}

The parent sample consists of 3,679 galaxies detected in HST images from three bands (F606W, F814W, and F160W). The specific process followed is outlined as follows: First, cutout images for each galaxy were generated with a size of $301 \times 301$ pixels ($~ 18.06^{\prime\prime} \times 18.06^{\prime\prime}$). Next, background noise estimation and subtraction were performed on each cutout image. Finally, the galaxy segmentation maps were generated using the \texttt{Image Segmentation} from \texttt{Photutils} \citep{Bradley+etal+2024}, with the detection threshold set to 1 $\sigma_{bkg}$. If a segmentation map was generated for a galaxy, it was considered detected by HST. To assess the detection capability of CSST, the same detection process was applied to CSST wide and deep field images across all 7 bands (NUV, u, g, r, i, z, and y). Cutout images of size $101 \times 101$ pixels ($~ 7.47^{\prime\prime} \times 7.47^{\prime\prime}$) were used, with the same detection threshold (1 $\sigma_{bkg}$) as in the HST analysis. As a result, 1,776 galaxies were detected in the wide-field images, and 3,363 galaxies were detected in the deep-field images.

For the non-parametric morphological measurements and comparisons, HST F814W images were used, along with CSST i-band wide and deep field images. The procedure for these measurements is as follows: First, cutout images were visually inspected to ensure no blended galaxies were present. Next, non-parametric morphological measurements were conducted using the code developed by \cite{Ren+etal+2023} for HST, CSST wide field, and deep field images. Measurements with an SNR lower than 2.5 were excluded, as these were considered unreliable \citep{Lotz+etal+2006}. Finally, galaxies with valid measurements in all three image types were selected. This resulted in a final sample of 289 galaxies, with redshifts in the range of $0.1 < z < 1.4$, magnitudes in the range of $17 < m_{\rm H} < 23$, and stellar masses in the range of $9.0 < \log\left(\rm M_{*} / M_{\odot}\right) < 11.4$.

\section{Results} \label{result}

\subsection{Capability of source detection}

\begin{figure*}[htbp]
    \centering
    \gridline{
        \fig{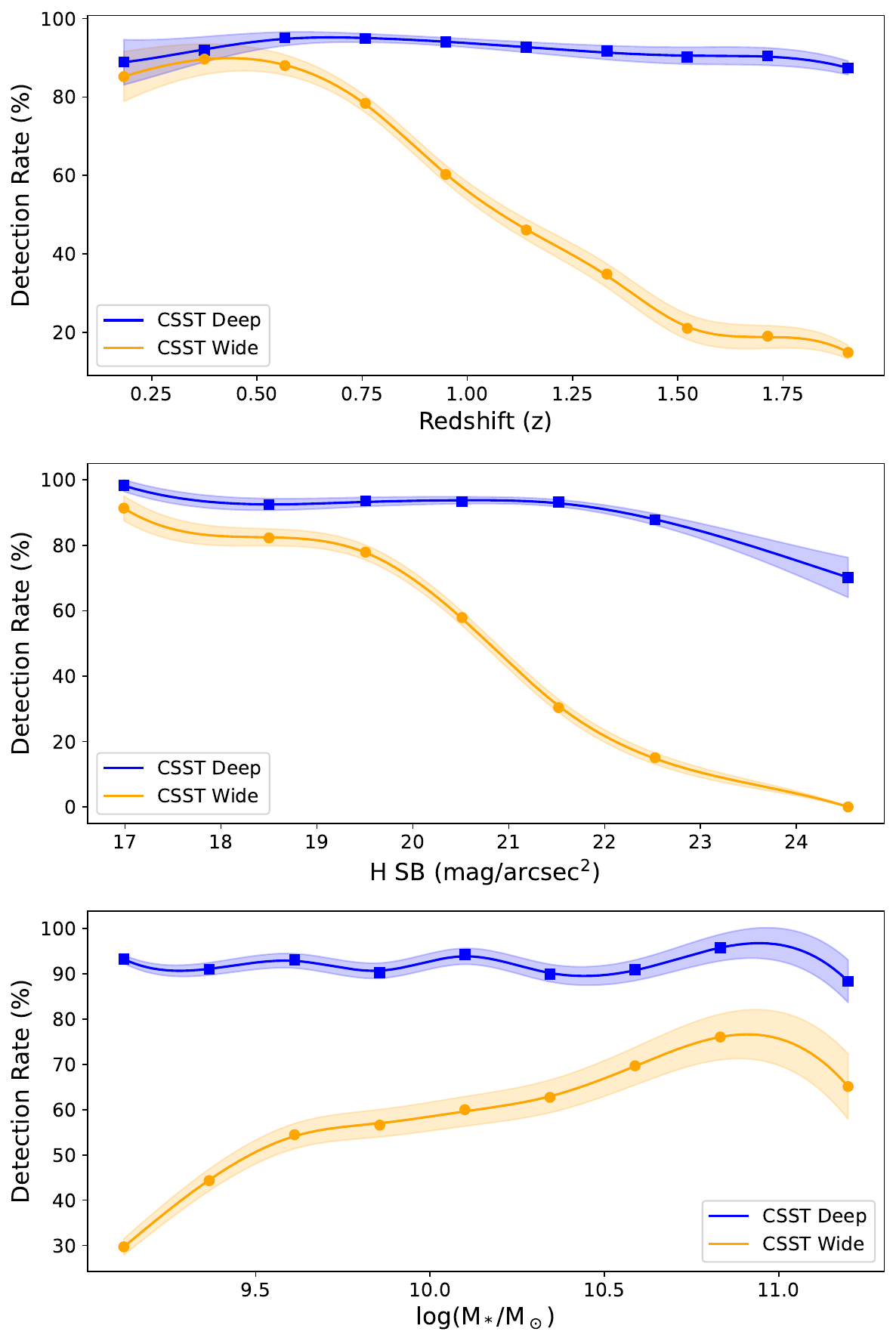}{0.48\textwidth}{(a)}
        \fig{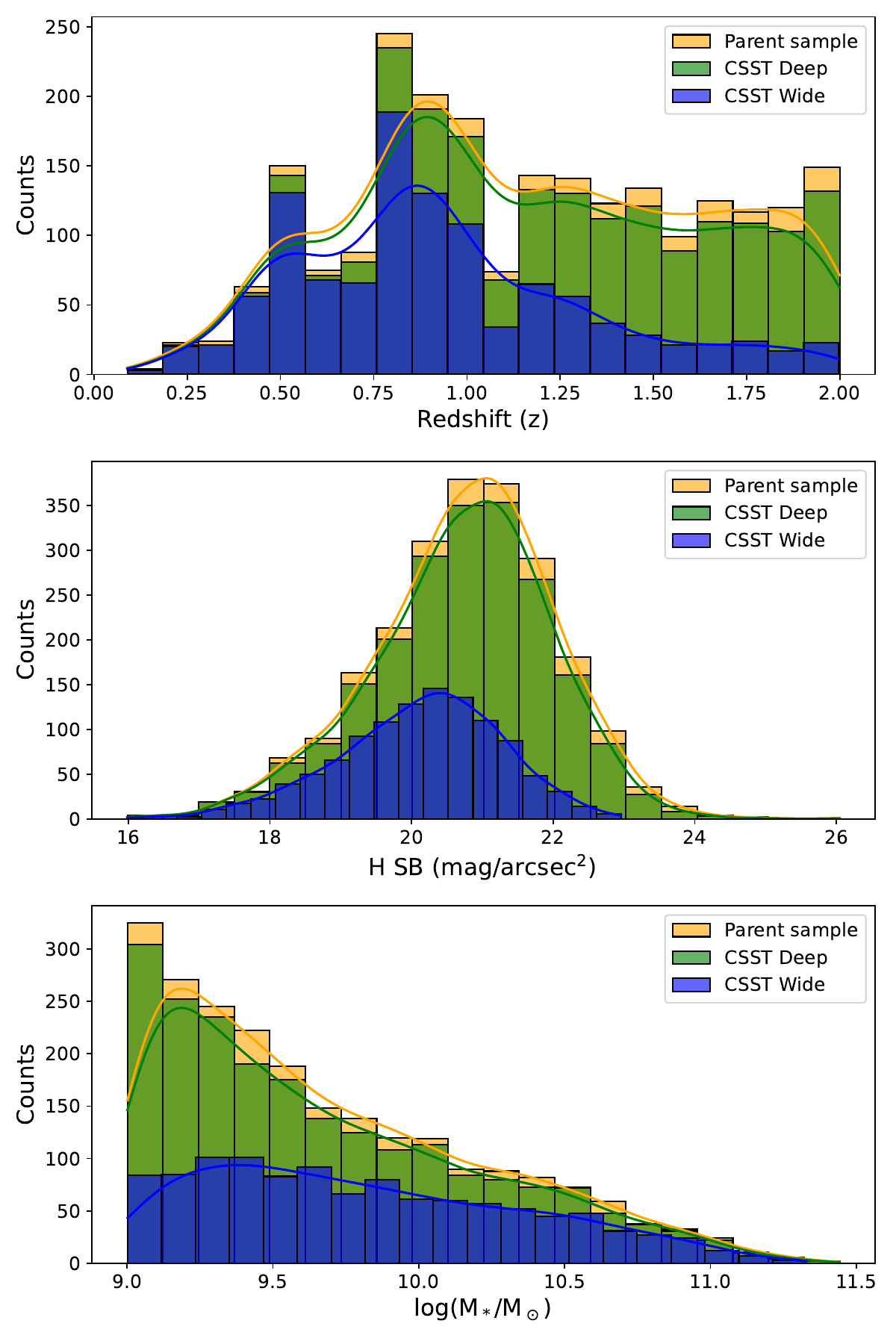}{0.48\textwidth}{(b)}
    }
    \caption{(a): Detection rate as a function of redshift, H band surface brightness, and stellar mass, with shaded regions representing the 68\% confidence interval.  
    (b): Distribution of redshift, H band surface brightness, and stellar mass, where yellow, green, and blue columns represent galaxies detected by HST images in 3 bands, CSST's deep-field images in 7 bands, and CSST's wide-field images in 7 bands.}
    \label{detection_comparison}
\end{figure*}

In this subsection, we estimate the detection capacity of CSST by counting the number of galaxies detected in the CSST wide and deep field images and comparing them with those detected by HST.

We prefer to use surface brightness rather than magnitude because galaxies are extended objects, and low-surface-brightness galaxies are more difficult to detect and more easily obscured by background noise than their high-surface-brightness counterparts. Therefore, surface brightness serves as a more suitable parameter for analysis.

In the left panel of Figure~\ref{detection_comparison}, we grouped the redshift, H band surface brightness, and stellar mass, and computed the detection rates for both the wide and deep fields within each group. The results show that the detection capacity of the CSST wide field attains an acceptable level for low-redshift, high-surface-brightness, and high-mass galaxies; however, it is significantly reduced when detecting high-redshift, low-surface-brightness, and low-mass galaxies. In the CSST deep field, the detection capacity is similar to that of HST deep field observations with cumulative exposure times greater than 10,000 seconds, except for a slight decline near to a surface brightness of 24 $\rm mag/arcsec^{2}$. Under such circumstances, the wide field is almost ineffective, while the deep field's detection capacity can still reach a level comparable to that of the HST.

\begin{figure*}[htbp]
	\centering
	\includegraphics[width=15cm, angle=0]{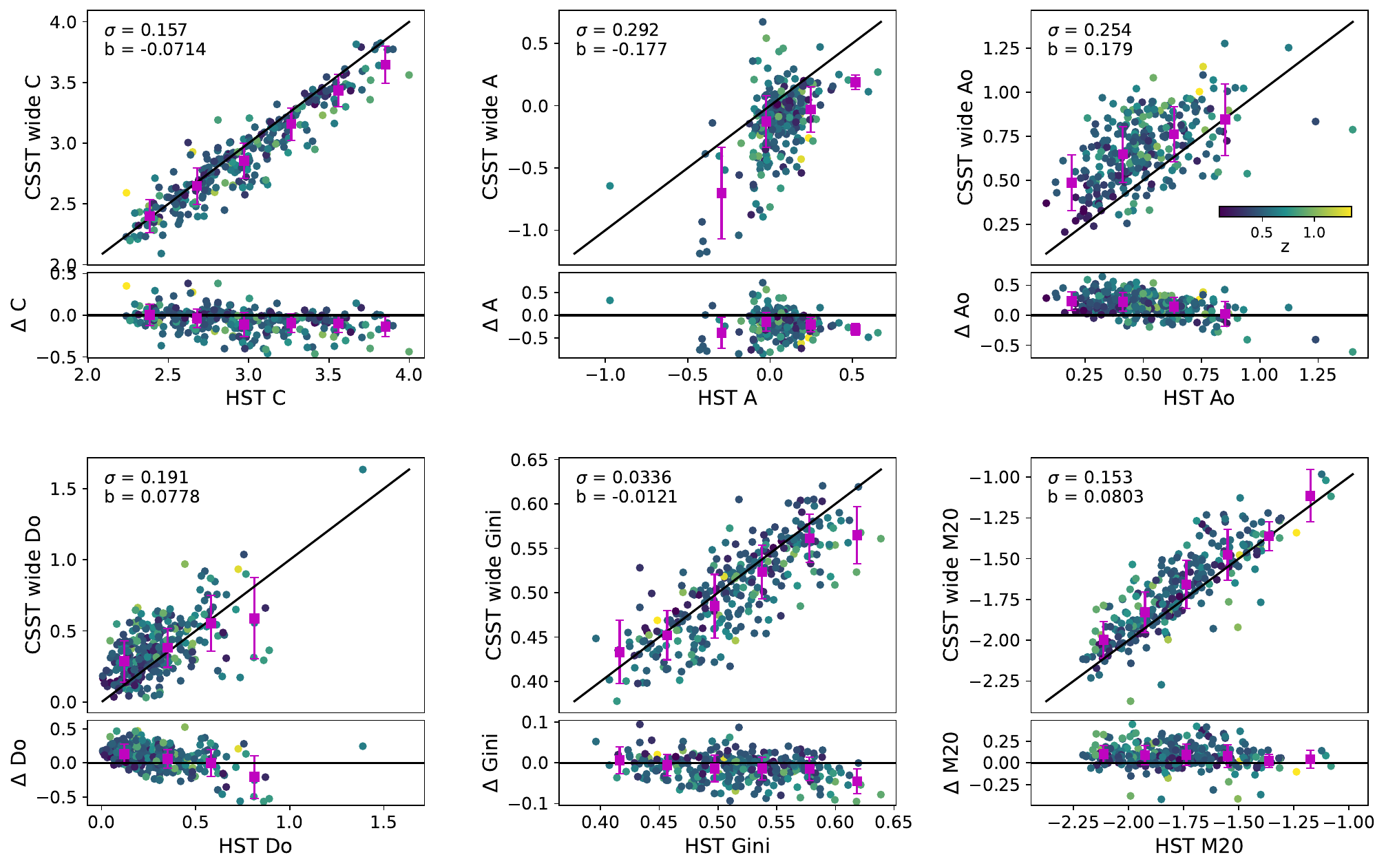}
	\caption{Comparison of the non-parametric morphological indicators ($C$,$A$,$A_{\rm O}$,$D_{\rm O}$,$Gini$,$M_{\rm 20}$) between galaxies observed by the HST and the corresponding wide field simulation images from the CSST. The color of the scatters represents the redshift. The blue points indicate the average data values within each bin, with error bars representing the scatter within 68\% of the samples in each bin.}
	\label{f814w_i_wide}
\end{figure*}

\begin{figure*}[htbp]
	\centering
	\includegraphics[width=15cm, angle=0]{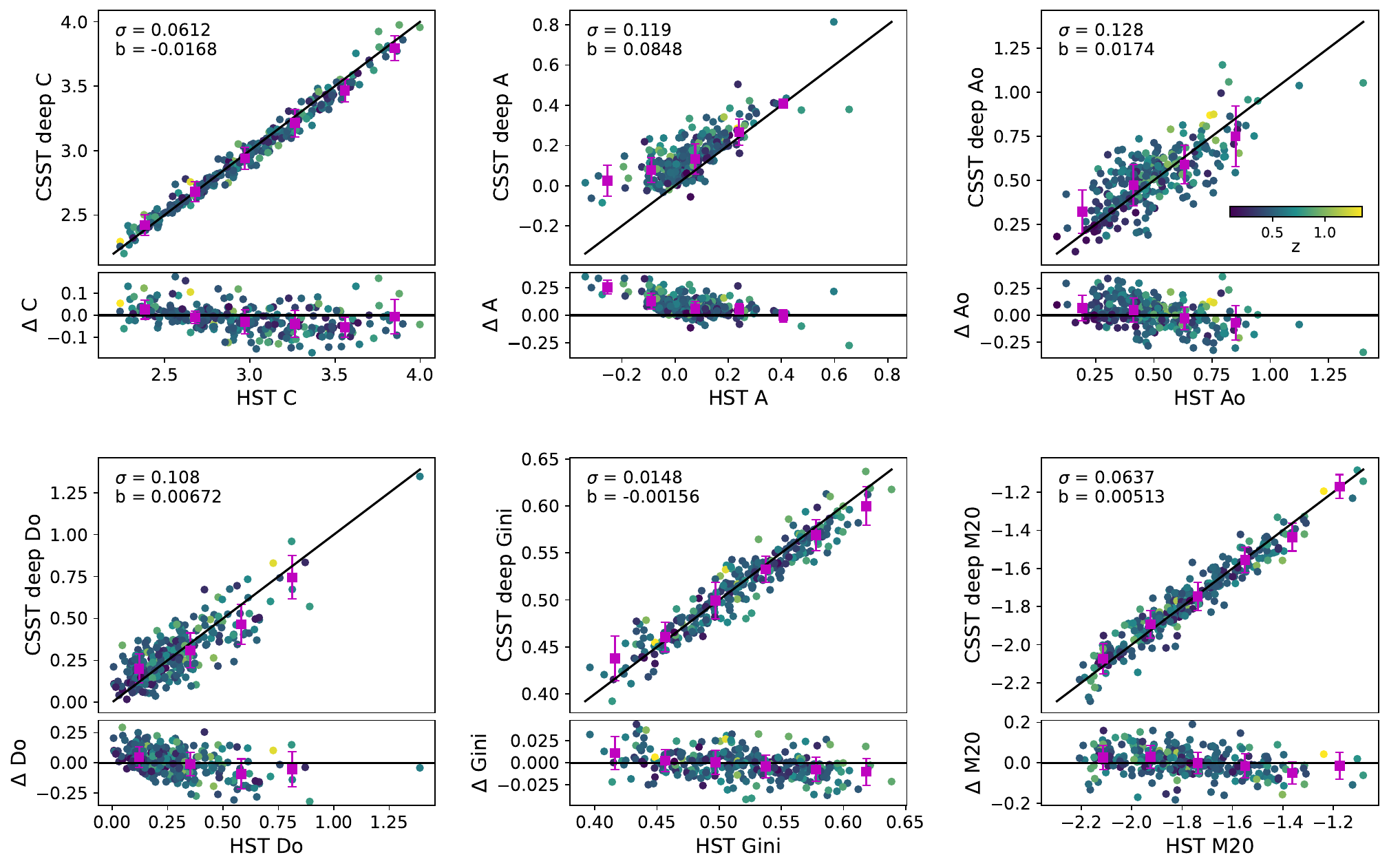}
	\caption{Comparison of the non-parametric morphological indicators ($C$,$A$,$A_{\rm O}$,$D_{\rm O}$,$Gini$,$M_{\rm 20}$) between galaxies observed by the HST and the corresponding deep field simulation images from the CSST. All markers are consistent with those used in Figure~\ref{f814w_i_wide}.}
	\label{f814w_i_deep}
\end{figure*}

The right panel of Figure~\ref {detection_comparison} presents the redshift, H band surface brightness, and stellar mass distributions of galaxies detected by HST, CSST's wide-field, and CSST's deep-field images. It is evident that the detection capacity of the CSST wide field is highly insufficient, especially for high-redshift, low-surface-brightness, and low-mass galaxies. In contrast, the detection capacity of the CSST deep field is comparable to that of the HST under various conditions.

Overall, when considering a complete sample, the detection capacity of the CSST deep field is satisfactory. Nevertheless, the detection capacity of the CSST wide field is lower than that of the deep field. Particularly for high-redshift, low-surface-brightness, and low-mass galaxies, its detection ability is inadequate and may cause biases.

\begin{figure}[htbp]
	\centering
	\includegraphics[width=8.2cm, angle=0]{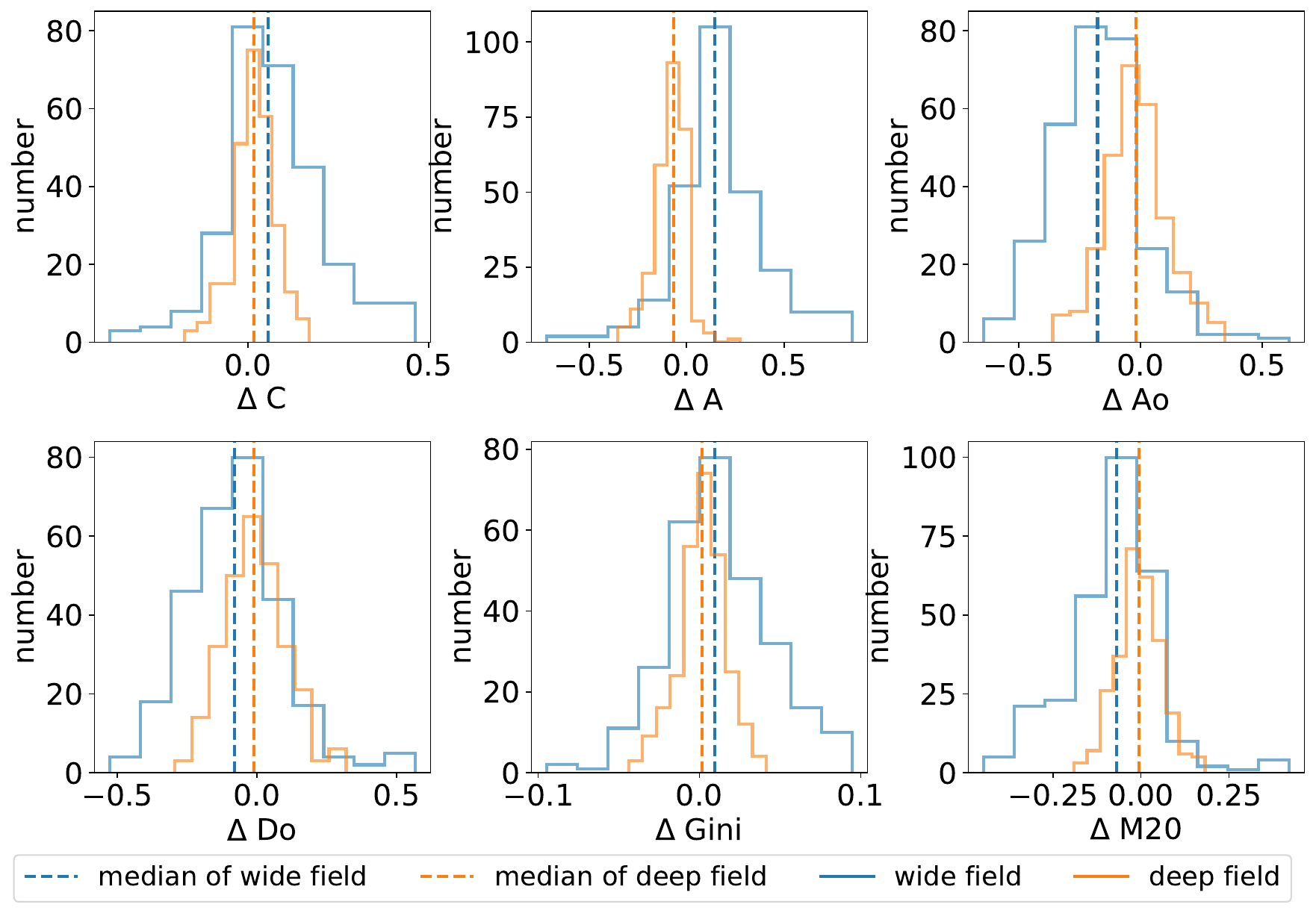}
	\caption{The distribution of differences between the morphological indicators of galaxies observed by the HST and their corresponding simulation images from the CSST.}
	\label{hist}
\end{figure}

\subsection{Accuracy of non-parametric morphological indicator measurements}

In this subsection, we investigated the accuracy of CSST wide and deep field measurements of galaxy non-parametric morphological indicators, with the HST measurements regarded as the truth. To reduce the influence of outliers, data points with CSST result errors greater than 3 times the $\sigma$ were removed iteratively. This operation will also be applied in the subsequent comparisons.

We studied the non-parametric morphological indicators of 289 galaxies, with their selection criteria provided in Section~\ref{measurements}. As can be seen from Figure~\ref{f814w_i_wide}, Figure~\ref{f814w_i_deep}, and Table~\ref{Tab_rb}, in the CSST's wide-field images, the measurements of $C$, $Gini$, and $M_{\rm 20}$ have basically reached the level of those derived from HST observations and have been further enhanced in the CSST deep field images, suggesting a better reconstruction of the central brighter structures in CSST images. We note that the $C$ values from CSST images are mostly slightly lower than those from HST images, which might be attributed to the larger FWHM of the point spread function (PSF) ($0.20^{\prime \prime}$ in the CSST simulated images and $0.10^{\prime \prime}$ in the HST images).

In CSST wide-field images, the consistency of $A$ is much lower when compared to HST images, but it improves in CSST deep-field images. This is because the SNR in the outer regions of galaxies is lower in CSST wide-field images, which leads to inconsistencies between the observed source shapes and the actual galaxy shapes. Most of the $A$ values from CSST's wide-field images are lower than those from HST, which may be ascribed to the higher noise level in CSST's wide-field images.

The $A_{\rm O}$ and $D_{\rm O}$ values measured from CSST's deep-field images are closer to those derived from HST images and show slightly better consistency compared to those from CSST's wide-field images. Nevertheless, there is still substantial scatter, indicating that the outer structure of CSST images is not well reconstructed or that these HST-based indicators have flaws when applied to CSST images. Since only CSST wide and deep field images were used here, the inconsistency between $A_{\rm O}$ and $D_{\rm O}$ may be due to the shallower depth of the survey. In the future, we will use CSST's ultra-deep-field images to measure the indicators and make comparisons.

Figure~\ref{hist} shows a comparative histogram of the differences between the measurement results of CSST and the true values. From this figure, it can be seen that, in CSST's deep-field images, the differences for all indicators ($C$, $A$, $A_{\rm O}$, $D_{\rm O}$, $Gini$, and $M_{\rm 20}$) are closer to zero compared to those from CSST's wide-field images, and their medians are also closer to zero. This suggests that the non-parametric morphological indicators ($C$, $A$, $A_{\rm O}$, $D_{\rm O}$, $Gini$, and $M_{\rm 20}$) from CSST's deep-field images are more consistent with the results obtained from HST than those from CSST's wide-field images.

\begin{figure}[htbp]
	\centering
	\includegraphics[width=8.5cm, angle=0]{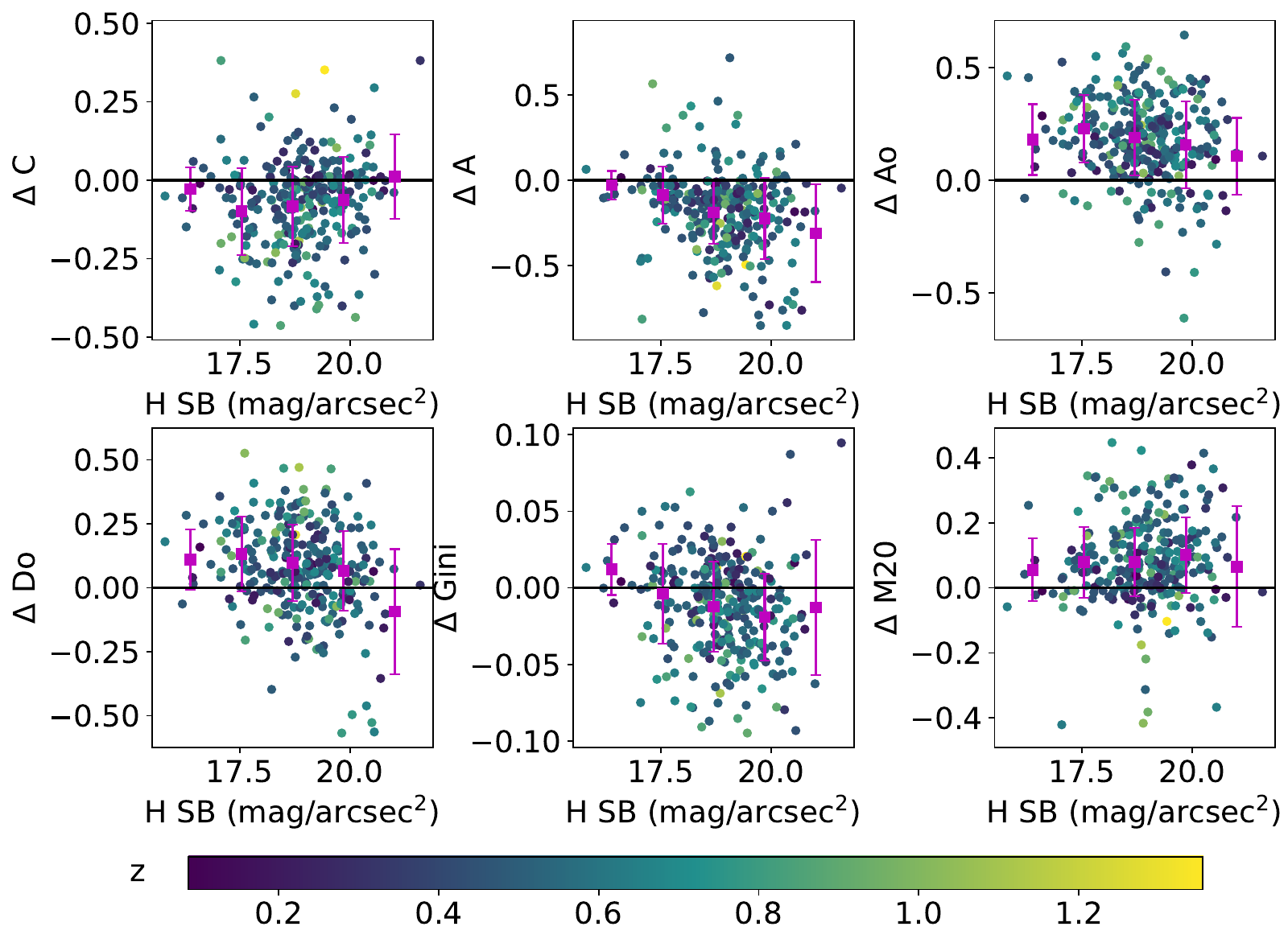}
	\caption{Relationship between the measurement differences of the CSST's wide-field images and the HST images and surface brightness. The blue points are consistent with those used in Figure~\ref{f814w_i_wide}.}
	\label{wfm}
\end{figure}

\begin{figure}[htbp]
	\centering
	\includegraphics[width=8.5cm, angle=0]{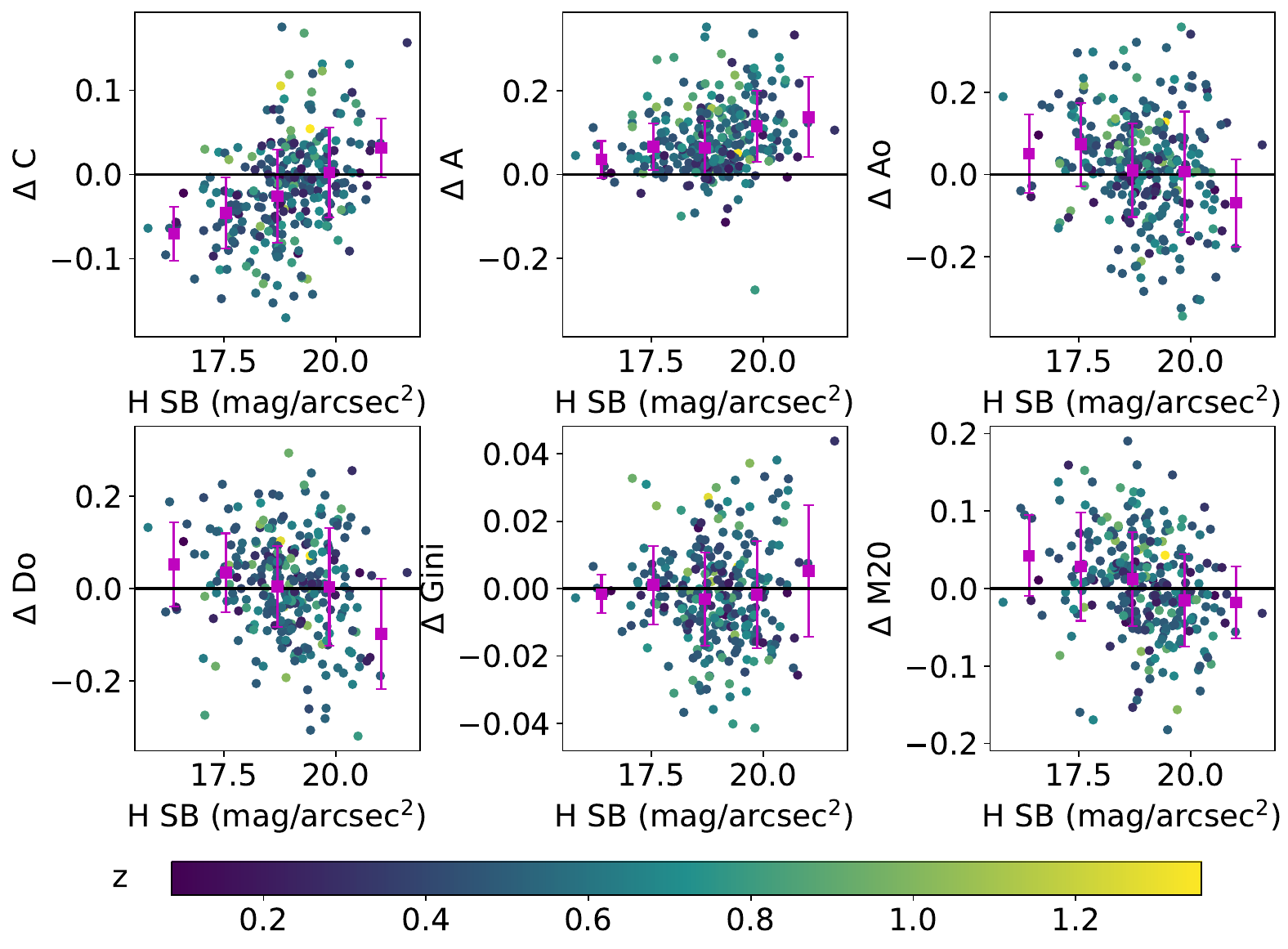}
	\caption{Relationship between the measurement differences of the CSST's deep-field images and the HST images and surface brightness. The blue points are consistent with those used in Figure~\ref{f814w_i_wide}.}
	\label{dfm}
\end{figure}

\renewcommand{\arraystretch}{0.95}
\begin{deluxetable*}{ccccc}
\tablenum{3}
\tablecaption{$\sigma$ and Bias of the CSST's wide-field and deep-field images.\label{Tab_rb}}
\tablewidth{0pt}
\tablehead{
\colhead{Indicator} & \colhead{Wide Field $\sigma$} & \colhead{Deep Field $\sigma$} & \colhead{Wide Field Bias} & \colhead{Deep Field Bias} \\[-15pt]
}
\startdata
$C$        & 0.16 & 0.06 & -0.07 & -0.02 \\  
$A$        & 0.29 & 0.12 & -0.18 & 0.08  \\
$A_{\rm O}$    & 0.25 & 0.13 & 0.18 & 0.02 \\  
$D_{\rm O}$    & 0.19 & 0.11 & 0.08 & 0.01 \\
$Gini$     & 0.03 & 0.01 & -0.01 & 0.00 \\  
$M_{\rm 20}$   & 0.15 & 0.06 & 0.08 & 0.01 \\
\enddata
\end{deluxetable*}

\subsection{Correction for Non-parametric Morphological Indicators from CSST}

Regarding the accuracy of non-parametric morphological indicators, it can be observed that the results obtained from CSST's wide-field images deviate considerably from HST results. However, for deep-field images, except for the asymmetric parameter $A$, the results can recover to a level close to those from HST. To make the CSST results more in line with those from HST, we propose an analytical correction method.

\begin{figure}[htbp]
	\centering
	\includegraphics[width=8.5cm, angle=0]{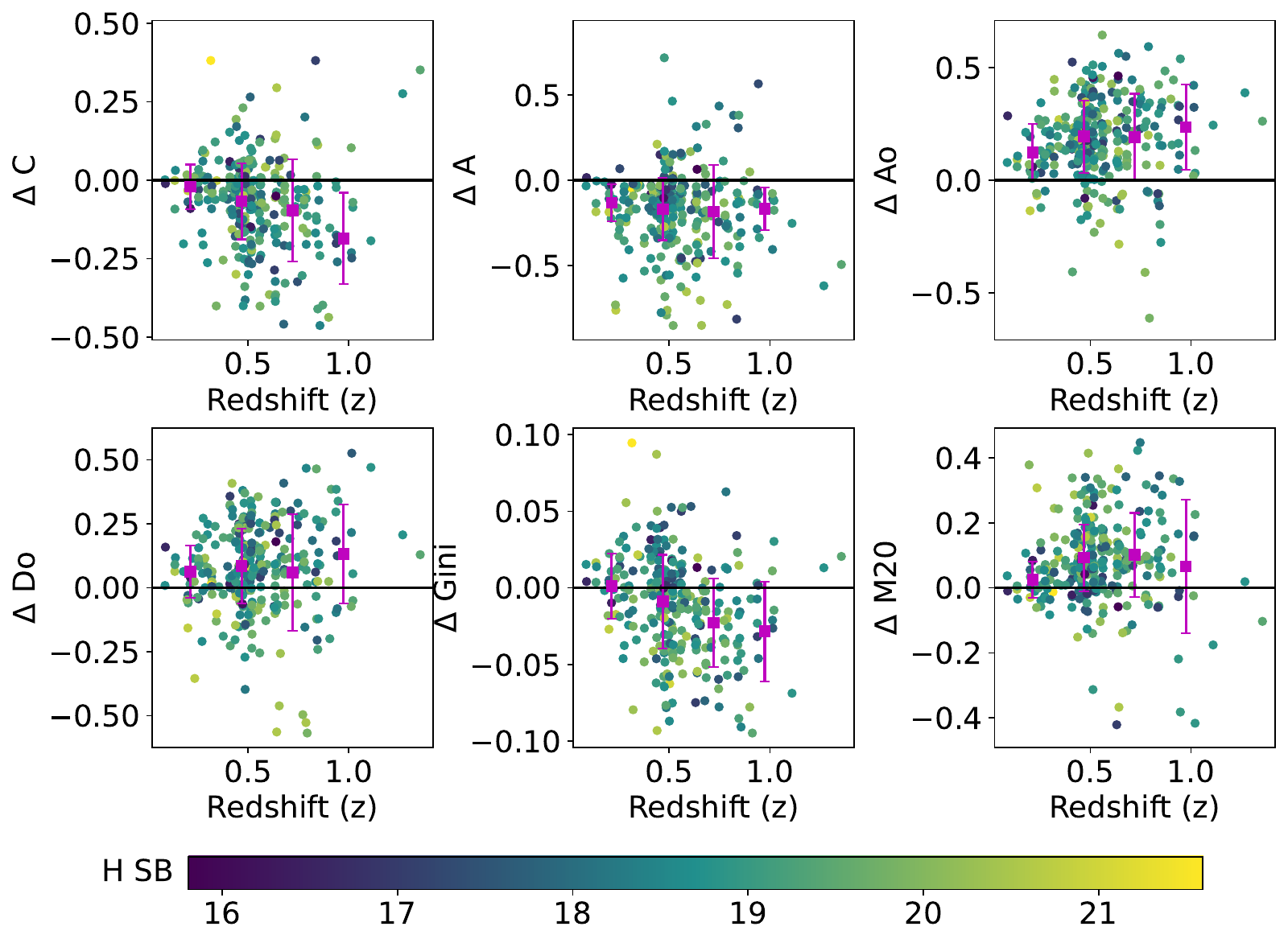}
	\caption{Relationship between the measurement differences of the CSST's wide-field images and the HST images and redshift. The blue points are consistent with those used in Figure~\ref{f814w_i_wide}.}
	\label{wfz}
\end{figure}

\begin{figure}[htbp]
	\centering
	\includegraphics[width=8.5cm, angle=0]{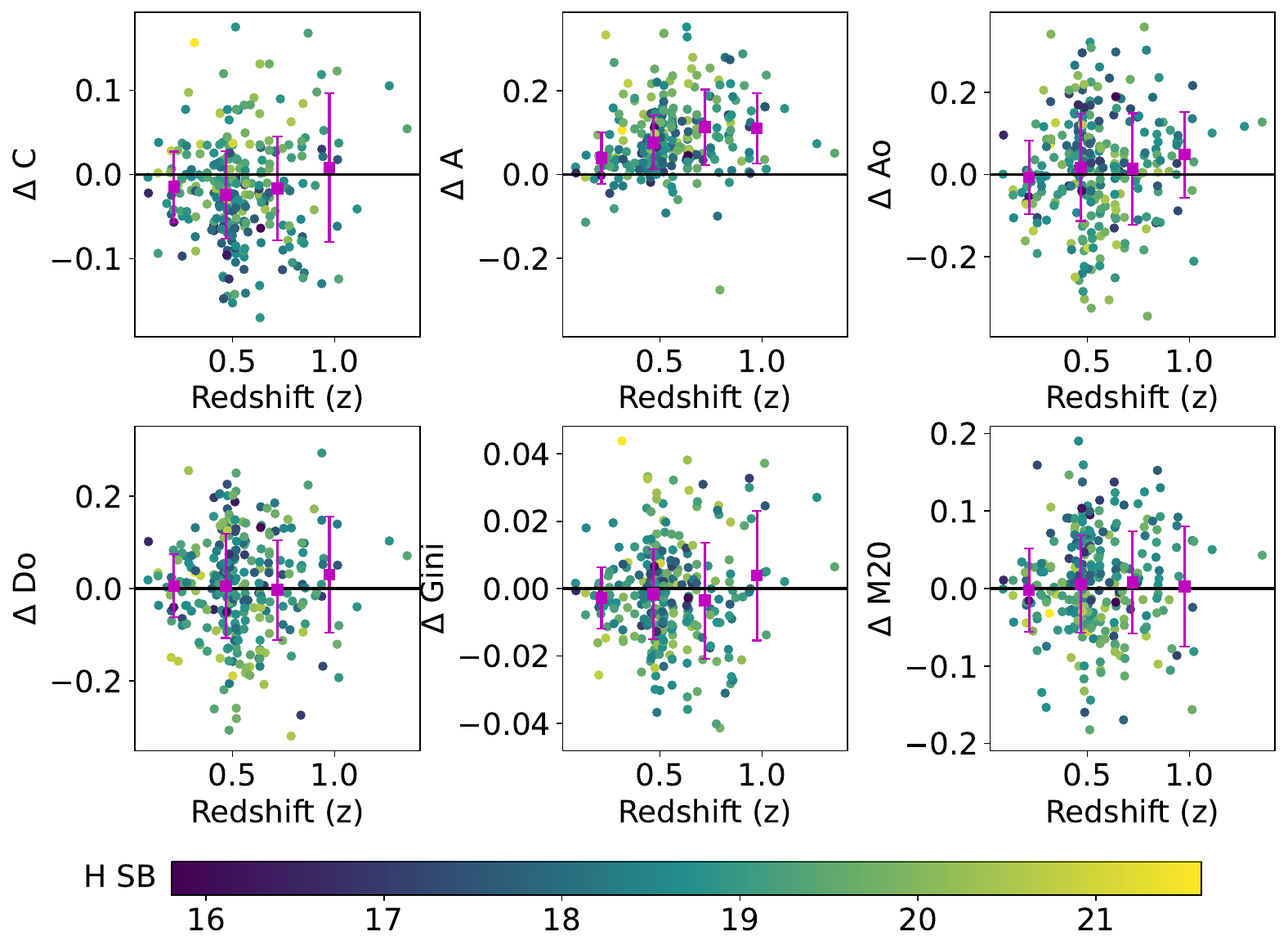}
	\caption{Relationship between the measurement differences of the CSST's deep-field images and the HST images and redshift. The blue points are consistent with those used in Figure~\ref{f814w_i_wide}.}
	\label{dfz}
\end{figure}

\begin{figure*}[htbp]
	\centering
	\includegraphics[width=15cm, angle=0]{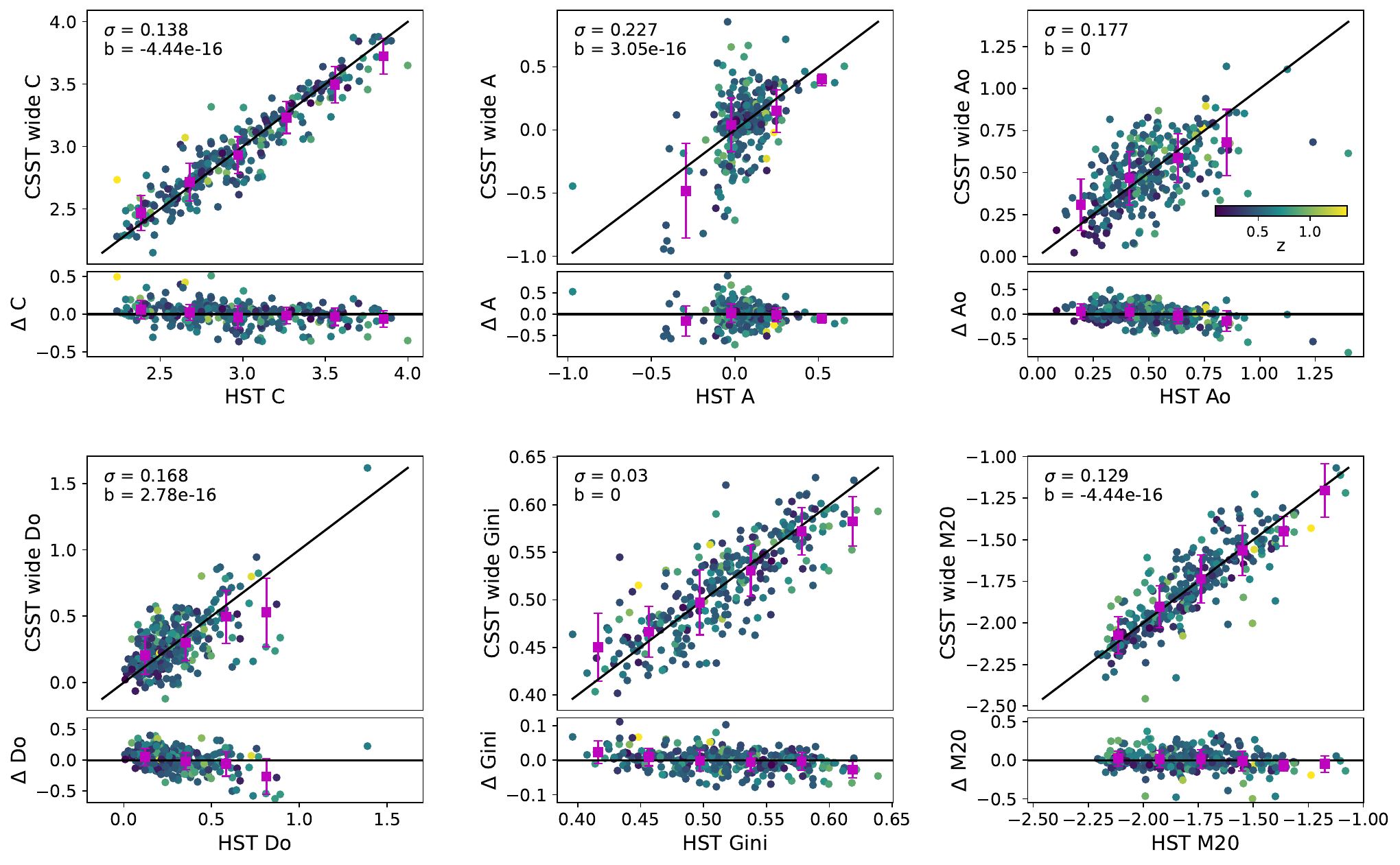}
	\caption{Comparison of the non-parametric morphological indicators ($C$,$A$,$A_{\rm O}$,$D_{\rm O}$,$Gini$,$M_{\rm 20}$) between galaxies observed by the HST and modified indicators of corresponding modified wide-field simulation images from the CSST. All markers are consistent with those used in Figure~\ref{f814w_i_wide}.}
	\label{wfmo}
\end{figure*}

\begin{figure*}[htbp]
	\centering
	\includegraphics[width=15cm, angle=0]{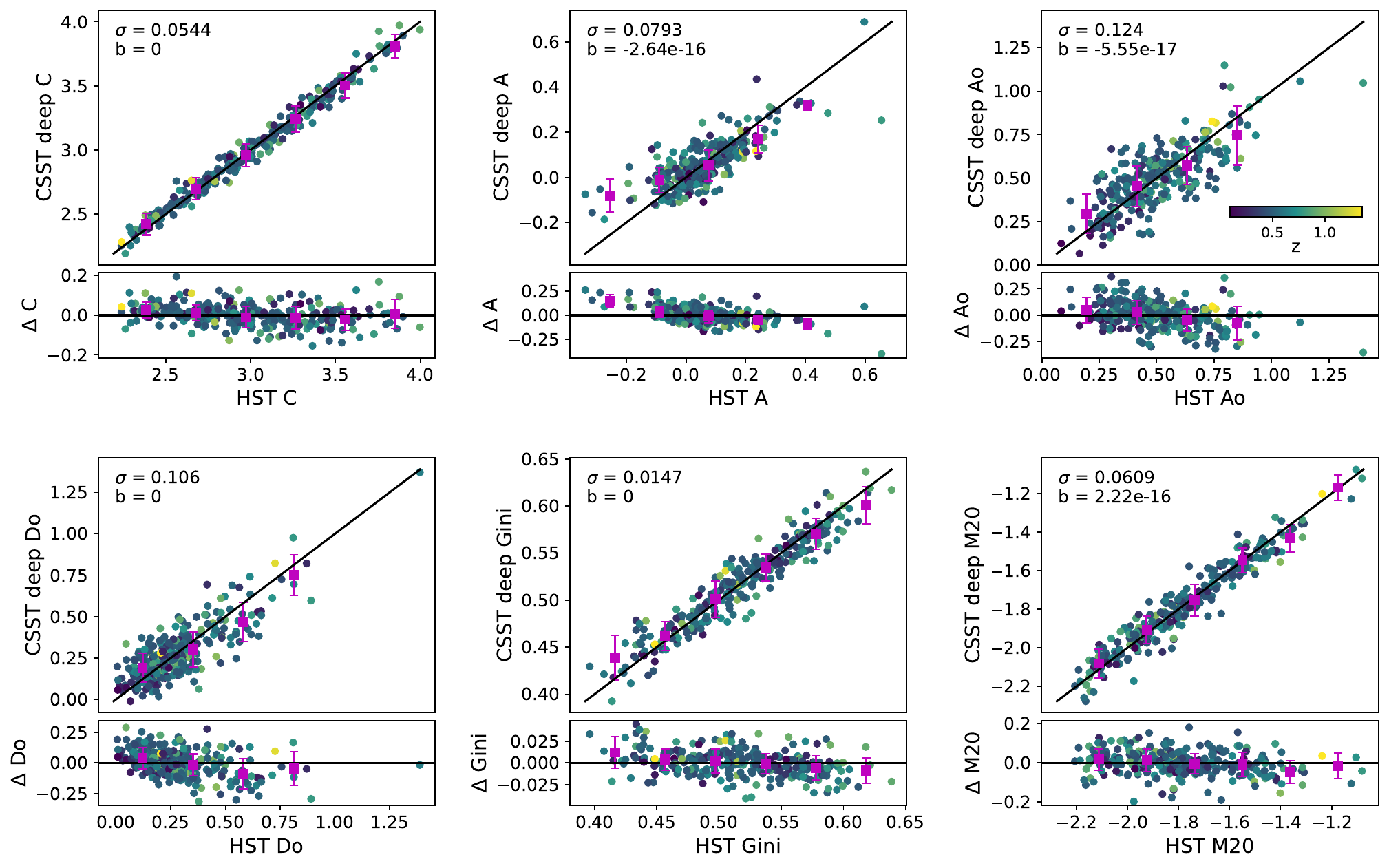}
	\caption{Comparison of the non-parametric morphological indicators ($C$,$A$,$A_{\rm O}$,$D_{\rm O}$,$Gini$,$M_{\rm 20}$) of galaxies observed by the HST and modified indicators of corresponding deep-field simulation images from the CSST. All markers are consistent with those used in Figure~\ref{f814w_i_wide}.}
	\label{dfmo}
\end{figure*}

Figure~\ref{wfm} and Figure~\ref{dfm}  show the relationship between the measurement differences of indicators and surface brightness. It can be seen that, in both CSST's wide-field and deep-field images, as the surface brightness increases, the scatter of various indicators gradually becomes more dispersed, and the length of the error bars (representing 1 $\sigma$) gradually increases. This indicates that the accuracy of measurements from CSST images (both wide-field and deep-field) decreases as the galaxy brightness weakens, and some indicators show systematic biases.

Figure~\ref{wfz} and Figure~\ref{dfz}  depict the relationship between the measurement differences of indicators and redshifts. As shown in the figures, as the redshift increases, some indicators gradually show systematic biases, and the scatter of the differences also increases. This implies that there is a certain dependence between the measurement differences and redshift.

\renewcommand{\arraystretch}{0.95}
\begin{deluxetable}{cccc}
\tablenum{4}
\tablecaption{Coefficients for the revised formula of morphological indicators in CSST's wide-field galaxy image.\label{Tab_cwf}}
\tablewidth{0pt}
\tablehead{
\colhead{Indicator} & \colhead{a} & \colhead{b} & \colhead{c} \\[-15pt]
}
\startdata
$C$        & -0.01 & 0.10 & 0.30 \\  
$A$        & 0.05  & 0.04 & -0.80 \\
$A_{\rm O}$    & 0.03  & -0.09 & -0.74 \\  
$D_{\rm O}$    & 0.04  & -0.07 & 0.86 \\
$Gini$     & 0.01  & 0.04 & -0.11 \\  
$M_{\rm 20}$   & -0.01 & 0.00 & 0.16 \\
\enddata
\end{deluxetable}

\renewcommand{\arraystretch}{0.95}
\begin{deluxetable}{cccc}
\tablenum{5}
\tablecaption{Coefficients for the revised formula of morphological indicators in CSST's deep-field galaxy image.\label{Tab_cdf}}
\tablewidth{0pt}
\tablehead{
\colhead{Indicator} & \colhead{a} & \colhead{b} & \colhead{c} \\[-15pt]
}
\startdata
$C$        & -0.02 & -0.02 & 0.47 \\  
$A$        & -0.02 & -0.09 & 0.35 \\
$A_{\rm O}$    &  0.03 & -0.05 & -0.50 \\  
$D_{\rm O}$    &  0.02 &  0.00 & -0.38 \\
$Gini$     &  0.00 &  0.00 & 0.01 \\  
$M_{\rm 20}$   &  0.02 & -0.01 & -0.35 \\
\enddata
\end{deluxetable}

To alleviate this problem, we have carried out a bivariate least-squares fit to model the relationship between the measurement differences, redshifts, and surface brightness, and subtracted them from the CSST image measurements, providing results for both wide and deep field observations. Indicators can be calibrated using the following formula ($P$ represents any indicator):

\begin{equation}
	P_{calibration}= P_{obs} + a*SB_{\rm H} +b*z + c
\end{equation}
	
In this revised formula, $P_{calibration}$ stands for the corrected indicators, $P_{obs}$ denotes the results measured from CSST wide or deep field images, $SB_{\rm H}$ represents H band surface brightness, and $z$ represents redshift.

Through iteration, the polynomial coefficients have been determined (shown in Table~\ref{Tab_cwf} and Table~\ref{Tab_cdf}). The comparison between the modified CSST image results and the HST image results is illustrated in Figure~\ref{wfmo} and Figure~\ref{dfmo}. It can be seen that, compared with the original Figures~\ref{f814w_i_wide} and~\ref{f814w_i_deep}, the systematic biases of each indicator have been corrected, and the scatter has slightly decreased. This method can be applied to the upcoming CSST images, especially for correcting the non-parametric morphological indicators measured in the survey images. It should be noted that the correction only deals with systematic biases and is not applicable to correcting individual galaxies.

\begin{figure}[htbp]
    \centering
    \includegraphics[width=8cm, angle=0]{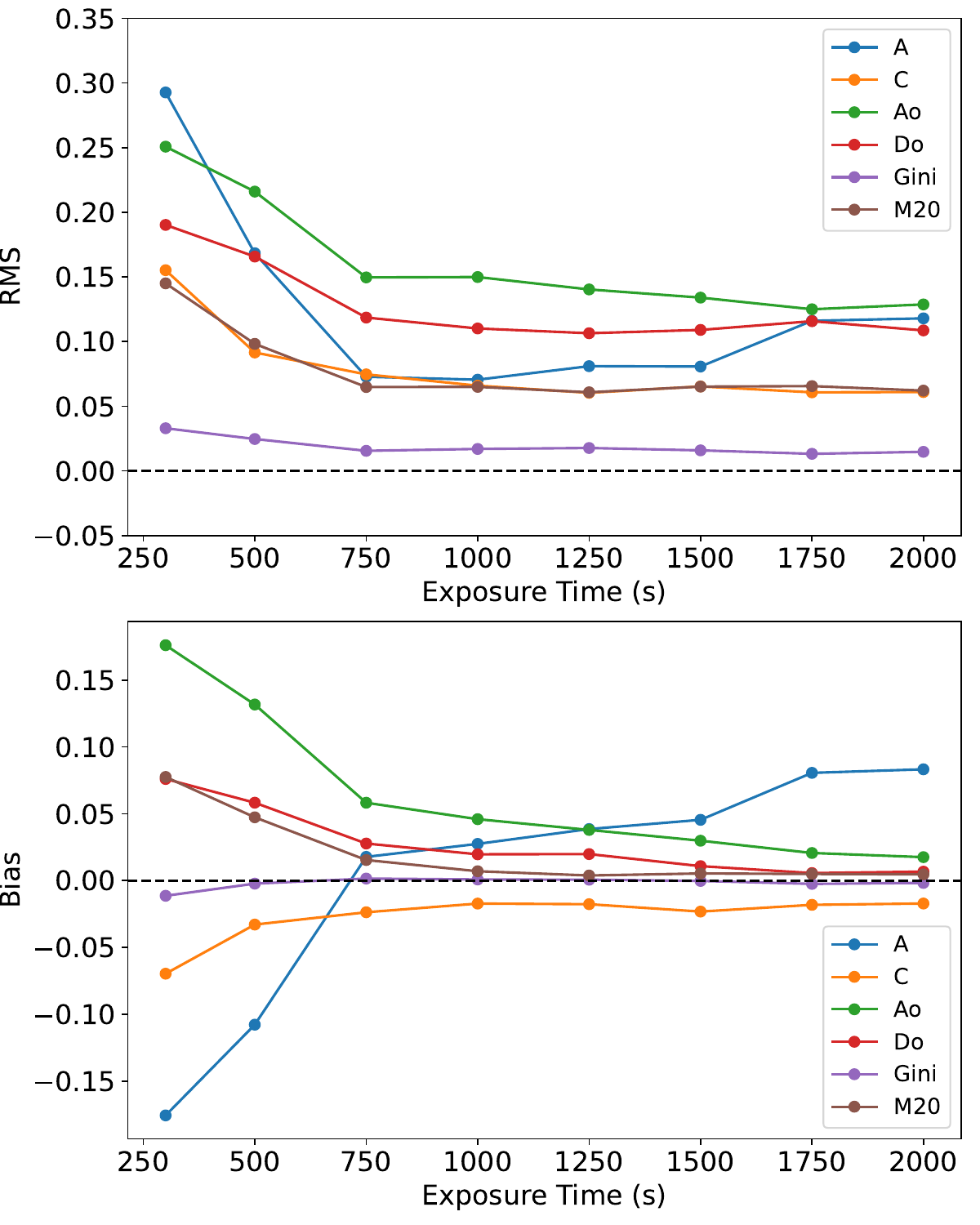}
    \caption{RMS and bias vary with exposure time. When exposure time reaches 750 seconds, the RMS and bias show only small changes with exposure time.}
    \label{rmse_bias_exposure_time}
\end{figure}

\begin{figure*}[htbp]
    \centering
    \includegraphics[width=15cm, angle=0]{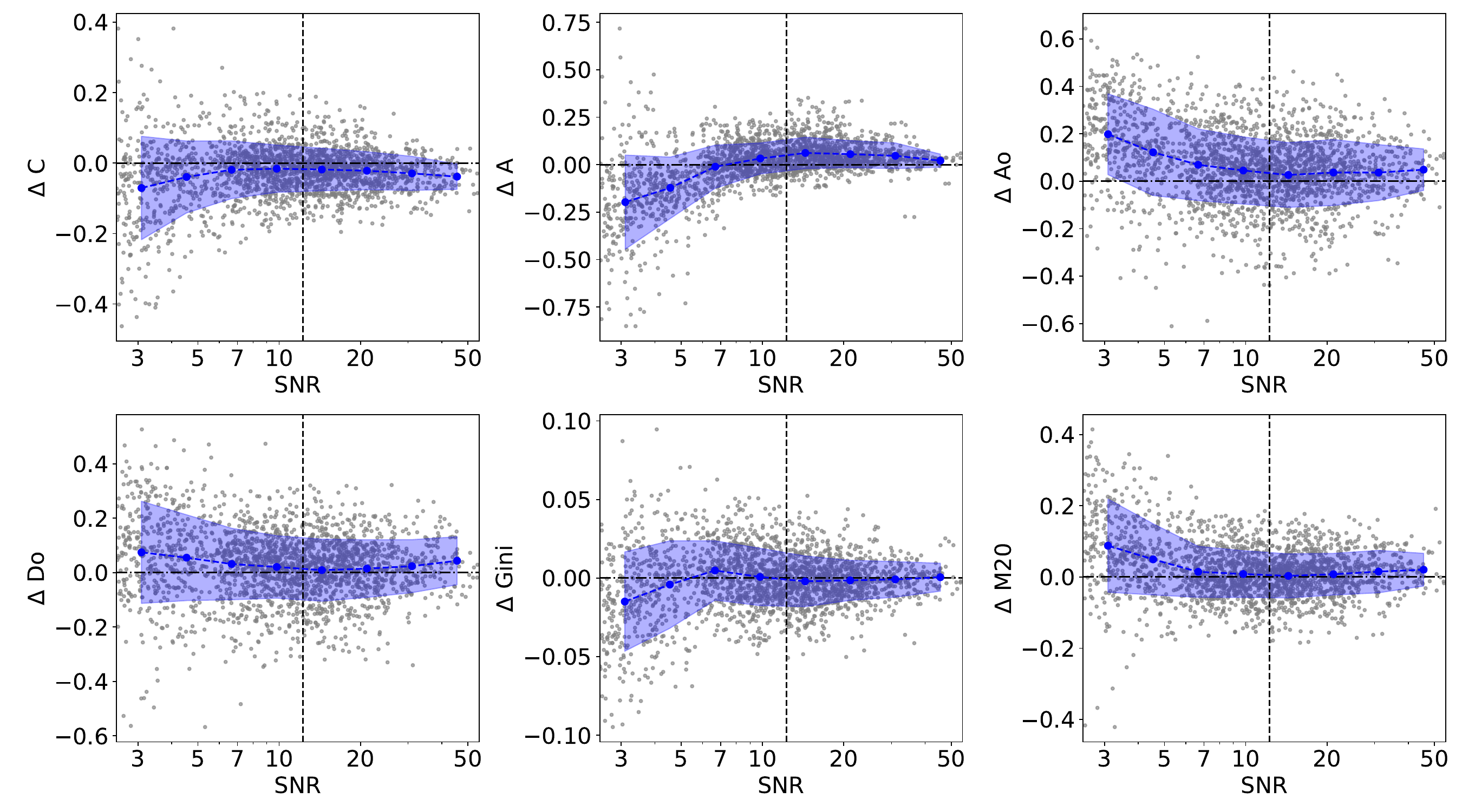}
    \caption{Difference change with SNR, with the blue shaded area representing the scatter within 68\% of the samples in each bin. The horizontal dashed line at zero represents perfect agreement between the measured values and those of HST, while the vertical dashed line marks the boundary at SNR=12.3, which is the average SNR for an exposure time of 750 seconds. Beyond this boundary, the consistency improves significantly.}
    \label{snr_difference}
\end{figure*}

\section{Discussions} \label{sect:discussion}

\subsection{The measurement accuracy of CSST depends on the exposure time}

The measurement accuracy is affected by the exposure depth. As depicted in Figure~\ref{f814w_i_wide} and Figure~\ref{f814w_i_deep}, the measurement accuracy in the CSST deep field has been greatly improved with the increase in exposure time. To assess the recovery of non-parametric indicators at different exposure depths, we measured the non-parametric morphological parameters of 289 galaxies in the CSST mock images, whose selection criteria are described in detail in Section~\ref{measurements}. The exposure times used were within the range of those for the wide-field and deep-field, specifically 500, 750, 1000, 1250, 1500, and 1750 seconds. After that, the measured results at different exposure times were compared with those from HST, and the RMS and bias were calculated.

RMS and bias represent the scatter and deviation between the CSST measurements and those from HST, reflecting the extent to which CSST recovers the non-parametric indicators. Figure~\ref{rmse_bias_exposure_time} shows the relationship between RMS, bias, and exposure time. It can be seen that, apart from $A$, as the exposure time increases, the RMS and bias of the other indicators ($C$, $A_{\rm O}$, $D_{\rm O}$, $Gini$, and $M_{\rm 20}$) gradually approach to zero. Moreover, the RMS and bias of each indicator show little change after the exposure time reaches 750 seconds. This implies that the accuracy of the non-parametric morphological measurements of the CSST mock images reaches a level comparable to that of the deep-field images when the exposure time reaches 750 seconds.

\subsection{The measurement accuracy of CSST depends on the signal-to-noise ratio}

The measurement accuracy may vary due to several reasons, and one of them is that the image SNR increases with exposure time. As shown in Figure~\ref{snr_difference}, the measurement difference of non-parametric morphological indicators ($C$, $A$, $A_{\rm O}$, $D_{\rm O}$, $Gini$, and $M_{\rm 20}$) between the CSST and the HST is presented as a function of SNR. It is observed that as the SNR rises, both the difference and the scatter decrease, suggesting that the CSST measurements are closer to those of HST. Among these indicators, those related to the light distribution in galaxies ($C$, $Gini$, and $M_{\rm 20}$) are all underestimated, while the ones describing the outer structure of galaxies ($A_{\rm O}$, $D_{\rm O}$) are overestimated. This might be because at a low SNR, the outer regions of galaxies are affected by noise, which further impacts the galaxy's segmentation map, as discussed by \cite{Wang+etal+2024}. The measurement results reach a stable and reliable level when the SNR is around 12.3, which is the average value of the image SNR when the exposure time is 750 seconds. This reflects the relationship between exposure time and measurement accuracy as shown in Figure~\ref{rmse_bias_exposure_time}, indicating that SNR is one of the factors influencing measurement accuracy.

Moreover, the improvement in measurement accuracy with increasing SNR also indirectly reflects the relationship between SNR and the surface brightness of galaxies. This relationship is predetermined during observations—galaxies with higher surface brightness naturally achieve higher SNR, leading to clearer structural details and better recovery of morphological indicators. Consequently, low surface brightness galaxies observed under low SNR conditions tend to exhibit larger deviations between their measured and the results of HST.

\section{Conclusions} \label{conclusion}

In this work, a volume-limited sample of 3,679 galaxies ($\log\left(M_{*} / M_{\odot}\right) > 9.0$ and $z < 2.0$) was constructed using GOODS-North HST observation images and corresponding CSST mock images. Subsequently, source detection was carried out on both types of images, and the non-parametric morphological parameters of galaxies were measured.

1. The galaxy detection capabilities of CSST's wide-field and deep-field observation modes were assessed. A complete sample was constructed using HST deep-field images. The CSST's wide-field can achieve detection capabilities comparable to the HST deep field observations under specific conditions (low-redshift, high-brightness, and high-mass). However, as redshift increases and surface brightness and stellar mass decrease, its detection ability deteriorates. In contrast, the CSST's deep-field performs better than the wide-field, especially for high-redshift, low-brightness, and low-mass galaxies, reaching detection capabilities similar to those of the HST deep field observations.

2. The accuracy of non-parametric morphological indicators for galaxies in CSST wide-and deep-field images was evaluated, with HST-based measurements as the true values. Measurements of $C$, and $Gini$, from both CSST fields are in line with HST-based results. The asymmetry indicator $A$ is closer to the results from HST in the CSST deep field compared to the wide field, yet biases still exist. For $A_{\rm O}$ and $D_{\rm O}$, the wide-field measurements deviate significantly from those of HST, while the deep-field measurements are more accurate. CSST's wide-field and deep-field images can be utilized to study bright substructures traced by $C$, $Gini$, and $M_{\rm 20}$ with high recovery. The indicators for detecting faint outer-region galaxy structures ($A_{\rm O}$ and $D_{\rm O}$) are applicable to CSST's deep-field images. When using the $A$ indicator to identify galaxy merger candidates, the measurement accuracy from CSST's deep-field images is not yet reliable, though CSST's ultra-deep-field images may perform better.

3. The difference between results measured by CSST and HST was found to be dependent on redshift and surface brightness. As the surface brightness of galaxies decreases, the scatter in differences increases and systematic bias emerges because of the decrease in SNR. By applying a revised formula, the systematic bias was calibrated, bringing CSST measurements closer to those from HST values. This enables the successful correction of non-parametric morphological indicators in the upcoming CSST survey images.

This work holds significant implications for future studies of the non-parametric morphological indicators of galaxies in CSST survey images. It provides a foundation and reference for further research in this area, helping to better understand the characteristics and properties of galaxies through CSST survey images.

\section*{acknowledgements}

The authors gratefully acknowledge the valuable feedback from the reviewers and the editorial team for enhancing the quality of this work. This project is supported by the National Natural Science Foundation of China (NSFC grants No. 12273052, 11733006, 12090040, 12090041, and 12073051) and the science research grants from the China Manned Space Project (No. CMS-CSST-2021-A04). NL acknowledges the support from the Ministry of Science and Technology of China (No. 2020SKA0110100), the science research grants from the China Manned Space Project (No. CMS-CSST-2021-A01), and the CAS Project for Young Scientists in Basic Research (No. YSBR-062).

\bibliography{PASPsample631}{}
\bibliographystyle{aasjournal}



\end{document}